\documentclass[aps,prd,twocolumn,times,graphicx,amssymb,tighten,showpacs,groupedaddress,superscriptaddress,floatfix]{revtex4-1}

\usepackage{graphicx,amssymb,amsmath}
\usepackage[dvips]{xcolor}
\usepackage{soul}
\usepackage{hyperref}
\newcommand{\be}{\begin{equation}}
\newcommand{\ee}{\end{equation}}

\begin{document}
\title{Charged-current inclusive neutrino cross sections in the SuperScaling model including quasielastic, pion production and meson-exchange contributions}

\author{M.V.~Ivanov}
\affiliation{Institute for Nuclear Research and Nuclear Energy, Bulgarian Academy of Sciences, Sofia 1784, Bulgaria}
\affiliation{Grupo de F\'{i}sica Nuclear, Departamento de F\'{i}sica At\'omica, Molecular y Nuclear, Facultad de Ciencias  F\'{i}sicas, Universidad Complutense de Madrid, Madrid E-28040, Spain}
\author{G.D.~Megias}
\affiliation{Departamento de F\'{i}sica At\'omica, Molecular y Nuclear, Universidad de Sevilla, 41080 Sevilla, Spain}
\author{R.~Gonz\'alez-Jim\'enez}
\affiliation{Department of Physics and Astronomy, Ghent University, Proeftuinstraat 86, B-9000 Gent, Belgium}
\author{O.~Moreno}
\affiliation{Center for Theoretical Physics, Laboratory for Nuclear Science and Department of Physics, Massachusetts Institute of Technology, Cambridge, Massachusetts 02139, USA}
\author{M.B.~Barbaro}
\affiliation{Dipartimento di Fisica, Universit\`{a} di Torino and INFN, Sezione di Torino, Via P. Giuria 1, 10125 Torino, Italy}
\author{J.A.~Caballero}
\affiliation{Departamento de F\'{i}sica At\'omica, Molecular y Nuclear, Universidad de Sevilla, 41080 Sevilla, Spain}
\author{T.W.~Donnelly}
\affiliation{Center for Theoretical Physics, Laboratory for Nuclear Science and Department of Physics, Massachusetts Institute of Technology, Cambridge, Massachusetts 02139, USA}

\date{\today}
\begin{abstract}
Charged current inclusive neutrino-nucleus cross sections are evaluated using the superscaling model for quasielastic scattering and its extension to the pion production region. The contribution of two-particle-two-hole vector meson-exchange current excitations is also considered within a fully relativistic model tested against electron scattering data. The results are compared with the inclusive neutrino-nucleus data from the T2K and SciBooNE experiments. For experiments where $\langle E_\nu \rangle \sim 0.8$~GeV, the three mechanisms considered in this work provide good agreement with the data. However, when the neutrino energy is larger, effects from beyond the $\Delta$ also appear to be playing a role. The results show that processes induced by two-body currents play a minor role at the kinematics considered.
\end{abstract}


\maketitle

\section{Introduction}
\label{intro}

New measurements of inclusive charged current (CC) neutrino-nucleus scattering cross sections, where only the outgoing lepton is detected, have been recently performed by the T2K~\cite{T2Kincl,T2Kincl_electron}, SciBooNE~\cite{SciBooNEincl} and ArgoNeuT~\cite{Argoneutincl1,Argoneutincl2} collaborations. For neutrino energies around $1$~GeV (T2K and SciBooNE) the main contributions to the cross sections are associated with quasielastic (QE) scattering and one pion (1$\pi$) production. These, along with the two-particle-two-hole (2p2h) meson-exchange current (MEC) contributions, are the only processes we shall consider in this paper, while at higher neutrino energies (ArgoNeuT) multiple pion and kaon production, excitation of resonances other than the $\Delta$ and deep inelastic channels should also be considered. That said, we do attempt to provide some insights into how important these last effects may become as the neutrino energy increases.

The QE muon neutrino and antineutrino cross sections measured by the MiniBooNE experiment~\cite{AguilarArevalo:2010zc,AguilarArevalo:2013hm}, where QE events are characterized by the absence of pions in the final state, have triggered a lot of theoretical work trying to explain the unexpectedly large results, in apparent tension with the higher-energy data from the NOMAD  experiment~\cite{Lyubushkin:2008pe}. Several calculations~\cite{Martini:2009aa, Amaro:2010sd, Nieves:2011yp, Lalakulich:2012ac} have demonstrated that 2p2h excitations induced by two-body meson-exchange currents (MEC) play a significant role in the interpretation of the QE MiniBooNE data and in the neutrino energy reconstruction, which is therefore model dependent. At a quantitative level, however, these calculations, relying on different models and approximations (see~\cite{Simo:2014wka} for a brief review of the various approaches), give quite different results. Furthermore, a model using Relativistic Green's Functions (RGF),
which does not explicitly contain two-body currents but to some extent includes  inelastic channels through a complex optical potential, has been shown to be able to explain the QE MiniBooNE data~\cite{Meucci}; however, these results depend significantly on the particular choice of the relativistic optical potential. Comparison with inclusive data, where many-particle (in particular two-nucleon) emission channels unambiguously contribute, can shed light on the role of MEC in neutrino and antineutrino scattering in different kinematical regions.

In this paper we evaluate the CC neutrino inclusive cross sections within the Superscaling approach (SuSA), introduced in~\cite{Amaro:2004bs} to describe neutrino-nucleus scattering by using electron scattering data instead of relying on specific nuclear models. This approach allows one to describe the QE and $\Delta$ resonance regions in a unified framework and can be applied to high energies due to its relativistic nature. In the QE region, the SuSA model has recently been improved in~\cite{Gonzalez-Jimenez:2014eqa} to incorporate effects arising in the Relativistic Mean Field (RMF) model in the  longitudinal and transverse nuclear responses, as well as in the isovector and isoscalar channels. Since MEC are known to violate superscaling, their contribution must be added to the superscaling result: this will be accomplished by using a parametrization~\cite{Megias:2014qva} of the relativistic calculation of \cite{De Pace:2003xu}. The parametrization is necessary in order to reduce the
computation time, since the exact calculation involves $7$-dimensional integrals.

The paper is organized as follows: in Sect.~\ref{model} we briefly describe the Superscaling approach to the QE and $\Delta$-resonance regions, including some recent improvements of the model. In Sect.~\ref{results} we test our approach with inclusive electron scattering data on $^{12}$C and we present the comparison of the calculation for muon-neutrino and electron-neutrino scattering with the data from the T2K and SciBooNE experiments; in the latter case we also present results for antineutrinos. In Sect.~\ref{Conclusions} we draw our conclusions.

\section{The SuperScaling model}\label{model}

The SuperScaling model, based on the superscaling properties of inclusive electron scattering~\cite{super,Chiara1}, has been extensively  used~\cite{Caballero:2005sj, Caballero:2006wi, Amaro:2006tf, Amaro:2006if, Gonzalez-Jimenez:2013plb} to predict neutrino and antineutrino cross sections for complex nuclei. The detailed description of the model can be found, {\it e.g.}, in \cite{Amaro:2004bs,Gonzalez-Jimenez:2014eqa}. Here we simply recall its main features. In the quasielastic peak (QEP) region the basic ingredient of the model is a phenomenological superscaling function
\be
f_L^{QE} = k_F \frac{R_L^{QE}}{G_L^{QE}}
\ee
extracted from the world electromagnetic $(e,e')$ data by dividing the {\it longitudinal} response $R_L^{QE}$ times the Fermi momentum $k_F$ by the single-nucleon elementary function $G_L^{QE}$.
The data show that $f_L^{QE}$ is to a large extent a function of only one variable, the scaling variable $\psi^\prime_{QE}$, and is independent of the momentum transfer $q$ (scaling of first kind) and of the nucleus, represented by the Fermi momentum $k_F$ (scaling of second kind).

The function $f_L^{QE}$ embeds most of the nuclear effects, both in the initial and in the final state, and can therefore be used to predict the weak charged current quasielastic (CCQE) $(\nu_l,l)$ cross section. In its  original version the SuSA model assumes that the superscaling function $f^{QE}$ is the same in the longitudinal ($L$) and transverse ($T$) channels, a property referred to as scaling of zeroth-kind.

The main merit of the SuSA model is the reasonable agreement, required by construction, with electron scattering data over a very wide range of kinematics and for a large variety of nuclei. Such an agreement is a crucial test for any nuclear model to be applied to neutrino reactions. Although phenomenological, the model has firm microscopic foundations in the RMF model, which is able to reproduce both the height and the asymmetric shape of the experimental superscaling function~\cite{Caballero:2007tz}. Furthermore, and importantly, the RMF model predicts a transverse superscaling function
\be
f_T^{QE} = k_F \frac{R_T^{QE}}{G_T^{QE}}
\ee
which is higher than the longitudinal one, a result supported by the separated $L/T$ data analysis~\cite{Finn:1984,super} and strictly linked to the relativistic nature of the  model~\cite{Ivanov:2013bta}. This result has recently been used to improve the ingredients of the SuSA model by constructing a new version (SuSAv2) where $f_T^{QE}>f_L^{QE}$~\cite{Gonzalez-Jimenez:2014eqa}. Moreover in SuSAv2 the effects of Pauli blocking, initially neglected, have been implemented. In the results we present in the next section the updated version SuSAv2 of the model will be used.

The scaling approach can then be inverted and predictions can be made for CCQE neutrino and antineutrino reactions by replacing the elementary electromagnetic vertex, $\gamma^*NN$, with the weak one, $WNN$.

In the QE region the superscaling predictions have been succesfully compared with the recent MINER$\nu$A data~\cite{Minerva1,Minerva2}, that have been shown in~\cite{Megias:2014kia, Gonzalez-Jimenez:2014eqa} to be well reproduced without need of invoking large 2p2h contributions. Good agreement is also obtained with the high-energy NOMAD data~\cite{Lyubushkin:2008pe}. On the contrary, the MiniBooNE QE data are underpredicted by the model. The inclusion of 2p2h MEC excitations in the vector channel, evaluated using the model of \cite{De Pace:2003xu}, gives results which are closer to the experimental points, but are not enough to explain the data~\cite{Amaro:2010sd,Amaro:2011aa}, unlike the analysis and results in~\cite{Martini:2011wp,Nieves:2011yp}.

It should be mentioned that the model developed in~\cite{De Pace:2003xu} for electron scattering only contains the vector part of the two-body current. Assuming the transverse vector 2p2h MEC scaling function, $f_{T,VV}^{MEC}$, to be equal to the axial-axial ($f_{T,AA}^{MEC}$) and vector-axial ($f_{T',VA}^{MEC}$) ones --- as considered in~\cite{Martini:2009aa} --- a final result in agreement with CCQE MiniBooNE data is found. However, such a result cannot be fully justified until a proper 2p2h MEC calculation for the axial-axial and vector-axial responses is completed. The full calculation, including the axial two-body current, is in progress~\cite{MEC-Quique} and once it is available it will allow us to test the quality of this approximation.

The superscaling approach has been extended from the QE domain into the region where the $\Delta$-excitation dominates. In \cite{Amaro:2004bs,Maieron:2009an} it has been shown that the residual strength in the resonance region, obtained by subtracting the QE contribution from the total cross section, can be accounted for by introducing a new scaling function $f^\Delta$ dominated by the $N\to\Delta$ and employing a new scaling variable, $\psi^\prime_\Delta$, which is suited to the resonance region. In this paper we revisit this approach by using the improved QE superscaling model, SuSAv2, and an updated parametrization of the 2p2h MEC response~\cite{Megias:2014qva}. This procedure yields a good representation of the electromagnetic response in both the QE and $\Delta$ regions, as we shall illustrate in the next section.

Two different parameterizations have been considered to deal with the less-known axial form factors $C^A_{3,4,5}$ appearing in the elementary $W^+ N \to \Delta^+$ transition current. One is taken from \cite{AlvarezRuso:1998hi} where the deuteron was studied and the other was introduced in~\cite{Paschos:2003qr}. The comparison of results obtained with the two parameterizations can be viewed as a measure of the degree of uncertainty that can be expected from the choice of the single-nucleon response for this reaction. In the present analysis, our results show a negligible dependence upon the choice of parametrization. Hence all results presented in this paper correspond to the model of \cite{AlvarezRuso:1998hi}.

The superscaling predictions for the Delta region have been compared in~\cite{Ivanov:2012fm} with the MiniBooNE data in the case of $\pi^+$ production for the $\nu_\mu$-CH$_2$ CC charged pion cross section~\cite{neutr_pi+}. The results obtained for the flux-averaged double- and single-differential cross sections as functions of the muon kinetic energy and scattering angle were found to be in good agreement with the data, whereas for the totally integrated unfolded cross section a somewhat different dependence on the neutrino energy was obtained from the one displayed by the data. It is also important to stress that the present scaling approach is expected to be valid only for those kinematical situations where the $\Delta$-resonance excitation is the dominant inelastic process. At higher energies heavier resonances can be excited and the deep inelastic scattering domain can be reached. In this case the phenomenological extension of the model developed in~\cite{Barbaro:2003ie}, based on direct fits to highly
inelastic $e-N$ scattering data, is more suitable to describe the inclusive cross section.

Finally, as first noticed in~\cite{Amaro:2004bs} and studied in more depth in~\cite{Maieron:2009an}, sizeable deviations ($10-15$\%) from scaling are observed in the region where the QE and $\Delta$ responses overlap. In this region effects stemming from correlations and 2p2h MEC can play an important role and they cannot be reproduced by models that assume impulsive, quasifree scattering on bound nucleons. Therefore these effects must be added to the QE and $\Delta$ scaling functions. This we do by using a parametrization~\cite{Megias:2014qva} of the results of De Pace {\it et al.}~\cite{De Pace:2003xu}, where
a fully relativistic calculation of the 2p2h MEC contribution to inclusive electron scattering was performed.

\section{Results}\label{results}

In this section we first set up the model and test it versus electron scattering $(e,e')$ data for the kinematics relevant for the present study. Then we apply it to the analysis of inclusive neutrino scattering and compare its predictions with the data taken by the T2K and SciBooNE collaborations.

\subsection{The non-quasielastic scaling function}\label{Delta}

In this subsection we construct a phenomenological scaling function to be used in the non-quasielastic (non-QE) region, assuming that this is dominated by the $\Delta$-resonance. We follow an updated version of the procedure described in~\cite{Maieron:2009an}. More specifically, we first define a non-QE experimental scaling function in this region, $f^{\rm non-QE}$. This entails subtracting from the $(e,e')$ double-differential cross section the SuSAv2 QE scaling predictions and the 2p2h MEC contribution given by~\cite{Megias:2014qva}:
\begin{eqnarray}
\left(\frac{d^2\sigma}{d\Omega d\omega}\right)^{\rm{non-QE}} &=&
\left(\frac{d^2\sigma}{d\Omega d\omega}\right)^{\rm exp} -
\left(\frac{d^2\sigma}{d\Omega d\omega}\right)_{\rm 1p1h}^{\rm QE, SuSAv2}
\nonumber\\
&-&
\left(\frac{d^2\sigma}{d\Omega d\omega}\right)_{\rm{2p2h}}^{\rm{MEC}} \,.
\end{eqnarray}
Then we define a superscaling function in the region of the $\Delta$ peak as follows:
\begin{equation}
f^{\rm non-QE}(\psi_\Delta) = k_F \frac{\left(\frac{d^2\sigma}{d\Omega d\omega}\right)^{\rm non-QE}}{\sigma_M (v_L G_L^\Delta + v_T G_T^\Delta)} \,,
\end{equation}
where $\psi_\Delta$ is the $\Delta$ scaling variable and $G_L^\Delta$, $G_T^\Delta$ are single-hadron functions referred to the $N\to\Delta$ transition (see~\cite{Maieron:2009an} for explicit expressions). In the above $\sigma_M$ is the Mott cross section and $v_{L,T}$ are the usual kinematic factors.

\begin{figure}[ht]\centering
\includegraphics[width=85mm]{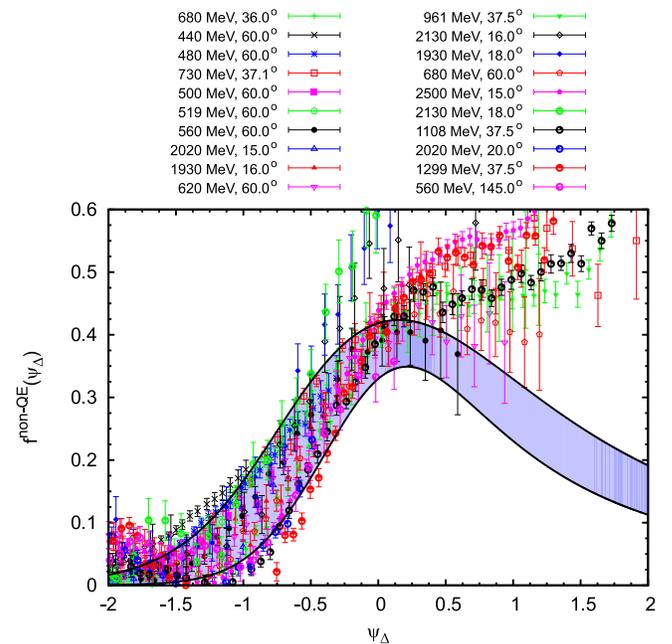}
\caption{(Color online) Averaged experimental values of
$f^{\rm non-QE}(\psi_{\Delta})$ together with a phenomenological fit of the non-QE scaling function. The colored band represents an estimation of the theoretical uncertainty (see text).
}
\label{fig:fDelta}
\end{figure}

\begin{figure*}[ht]
\centering
  \includegraphics[width=80mm]{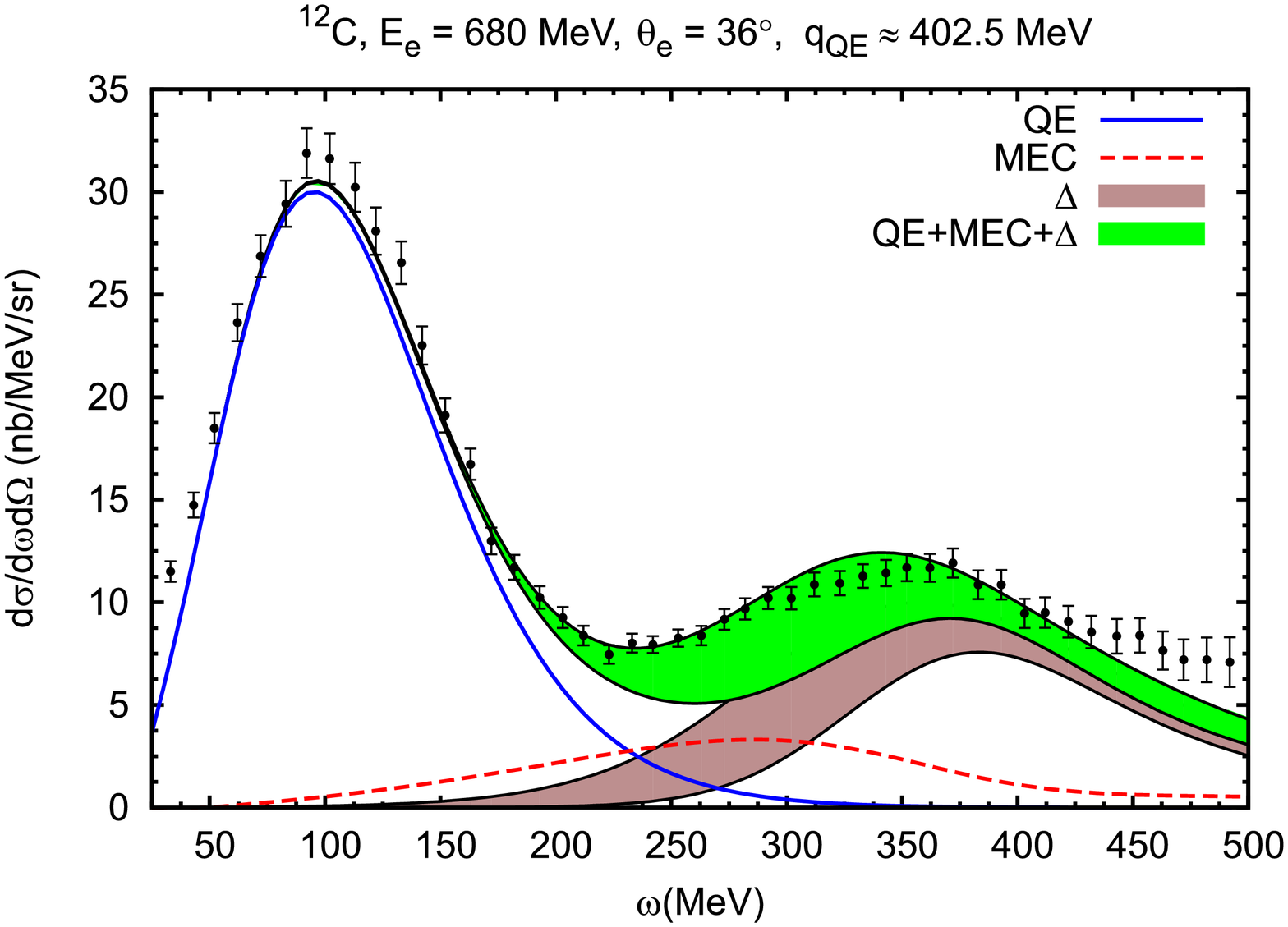} \hspace{0.5cm}
  \includegraphics[width=80mm]{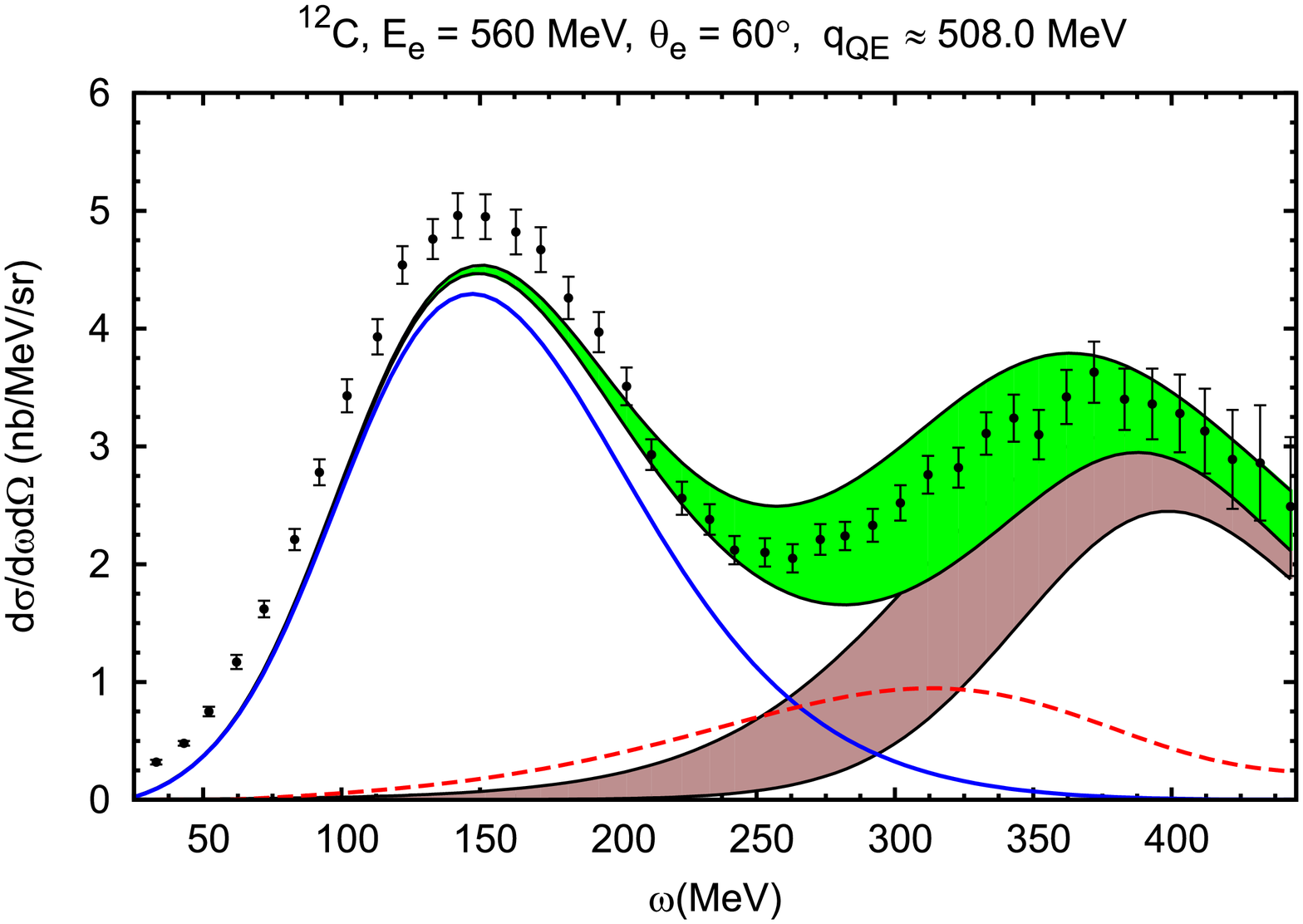} \\[3pt]
  \includegraphics[width=80mm]{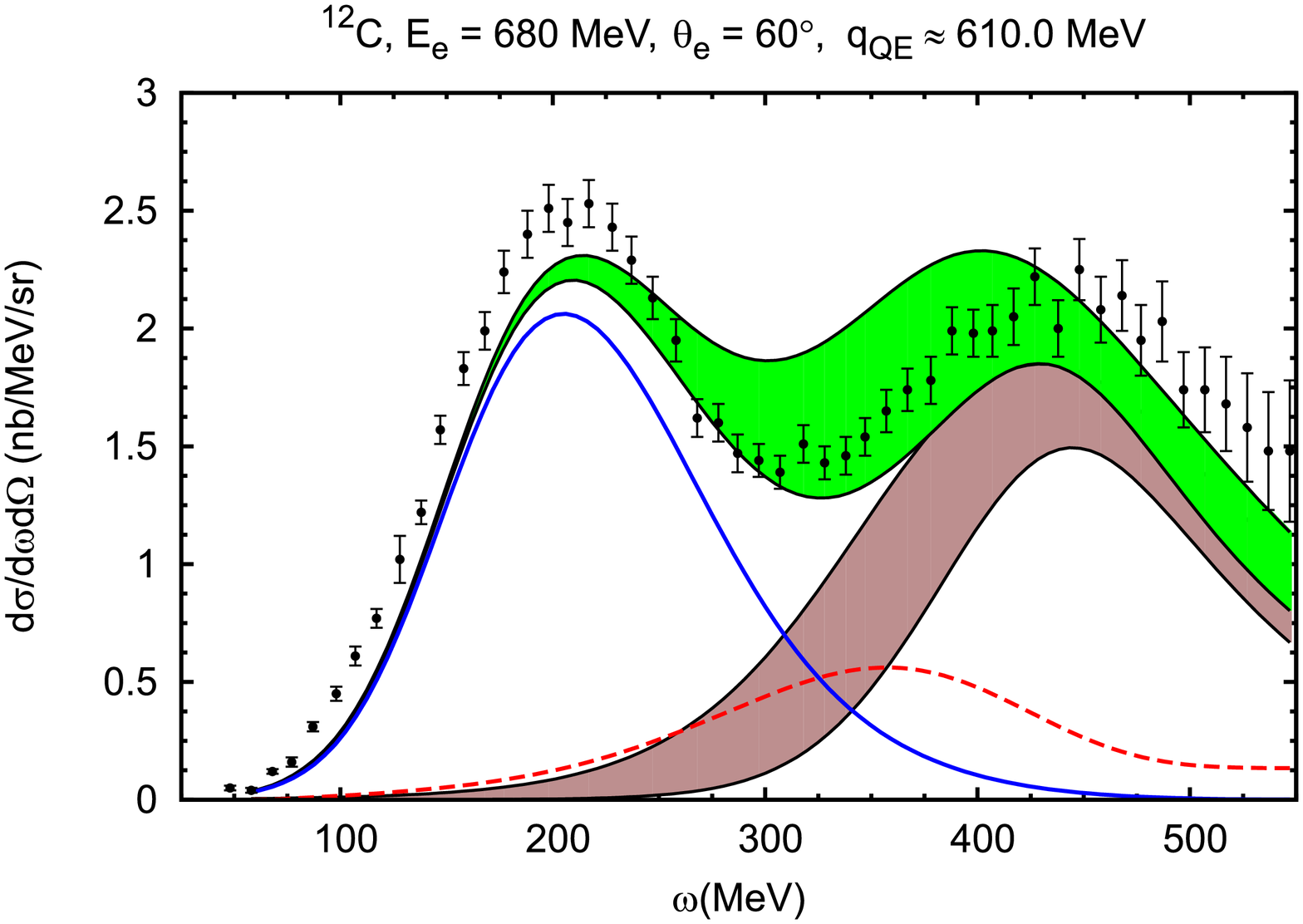} \hspace{0.5cm}
  \includegraphics[width=80mm]{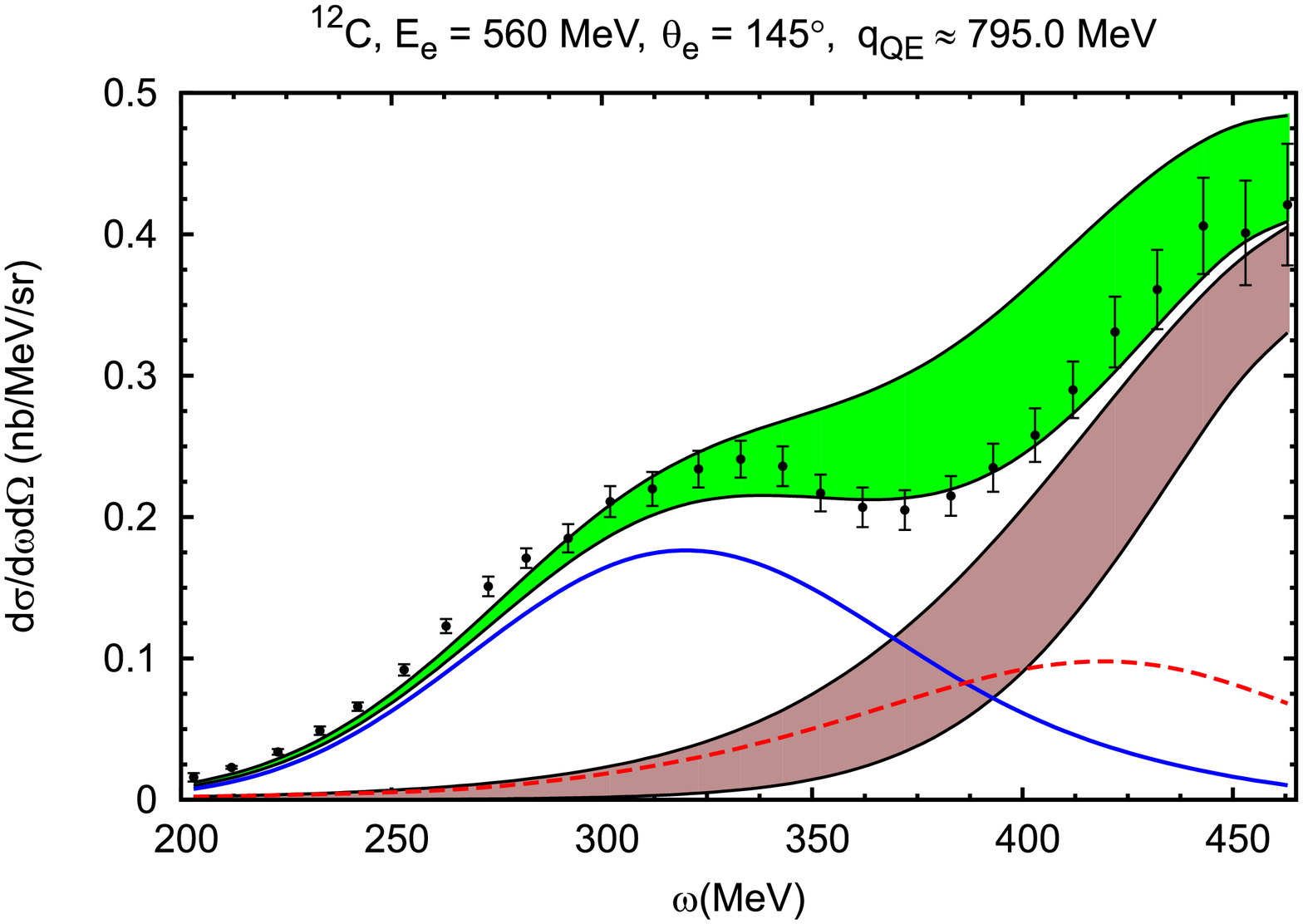}
\caption{Double-differential inclusive electron-carbon cross sections, $d\sigma/d\omega d\Omega$. The panels are labeled according to beam energy,
scattering angle, and value of $q_{\text{\tiny QE}}$ at the quasielastic peak. The results are compared with the experimental data from \cite{QESarchive} at the selected kinematics (see text).}
\label{fig:test}
\end{figure*}

Next we study the scaling behavior of $f^{\rm non-QE}$ by analyzing a large set of high quality $(e,e')$ data for $^{12}$C, using similar procedures to those discussed in \cite{Amaro:2004bs}. The data used there (see also~\cite{QESarchive}) were chosen to match --- at least roughly --- the kinematics that are relevant for the neutrino data under discussion; these choices of kinematics are listed in the figure.

From this analysis, illustrated in Fig.~\ref{fig:fDelta}, it appears that scaling in the $\Delta$ region works reasonably well up to the center of the $\Delta$ peak, $\psi_\Delta=0$, while it breaks, as expected, at higher energies where other inelastic processes come into play. However the quality of scaling is not as good as in the QEP region. For this reason the non-QE scaling function is represented with a band, rather than with a function, which accounts for the spread of pseudo-data seen in Fig.~\ref{fig:fDelta}. This band, together with the SuSAv2 phenomenological fits and the MEC response, can now be used to test the model against electron scattering data and to predict neutrino and antineutrino cross sections.

\subsection{Test versus electron scattering}\label{eeprime}

In Fig.~\ref{fig:test} we compare the model predictions with inclusive electron scattering data on $^{12}$C. Although many high quality electron scattering data exist, here we only show results for a few representative choices of kinematics, similar to those involved in the neutrino experiments that we address in the following sections.

As observed, the model gives a good description of the data, provided the 2p2h MEC are included. These, as expected, play a major role in filling the ``dip'' region between the QE and $\Delta$ peaks. The band in the final cross section (green region) comes from the uncertainty in the determination of the $\Delta$ superscaling function. This explains that the data located in the region close to the $\Delta$-peak are contained within the limits of the above band.
More importantly, the model (with its associated uncertainty) is capable of reproducing successfully all data with particular emphasis on the dip region. This result gives us confidence in the reliability of the model and its application to the analysis of neutrino-nucleus scattering reactions.

\begin{figure*}[ht]
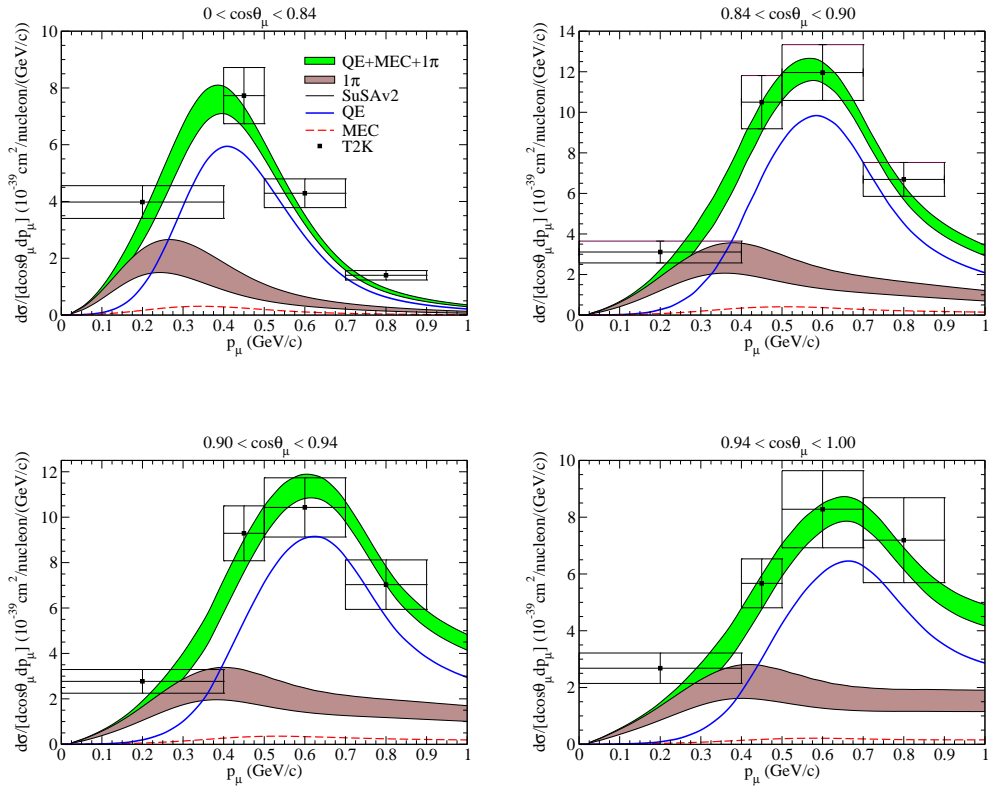

  \includegraphics[scale=0.25]{t2k0084_poster.eps} \hspace{0.5cm}
  \includegraphics[scale=0.25]{t2k084090_poster.eps} \\
\vspace{1.cm}
  \includegraphics[scale=0.25]{t2k090094_poster.eps} \hspace{0.5cm}
  \includegraphics[scale=0.25]{t2k094100_poster.eps}
\caption{(Color online) The  CC-inclusive T2K flux-folded $\nu_\mu$-$^{12}$C double-differential cross section per nucleon evaluated in the SuSAv2 model is displayed as a function of the muon momentum for different bins in the muon angle. The separate contributions of the QE, 1$\pi$ and vector 2p2h MEC are displayed. The data are from \cite{T2Kincl}.}
\label{fig:T2K}
\end{figure*}

\begin{figure}[ht]
  \includegraphics[width=80mm]{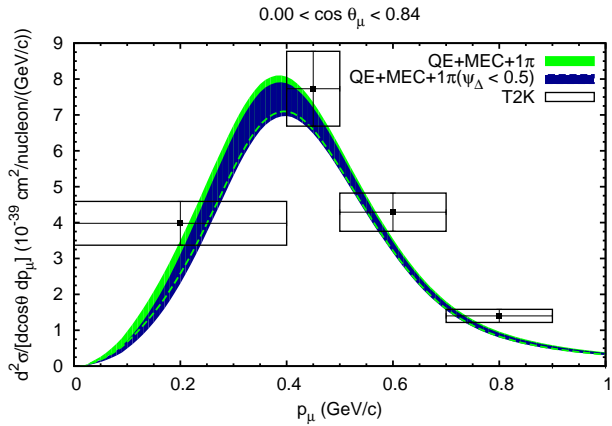}
  \caption{(Color online) The CC-inclusive T2K flux-folded $\nu_\mu$-$^{12}$C double-differential cross section per nucleon is displayed as a function of the muon momentum, which  corresponds to a bin in muon angle $0.00 < \cos\theta_\mu<0.84$. The full results with [QE+MEC+1$\pi$] and without [QE+MEC+1$\pi$($\psi_\Delta<0.5$)] the high-energy tail are shown. The data are from \cite{T2Kincl}.}\label{fig:T2K_cuts}
\end{figure}

\begin{figure*}[ht]
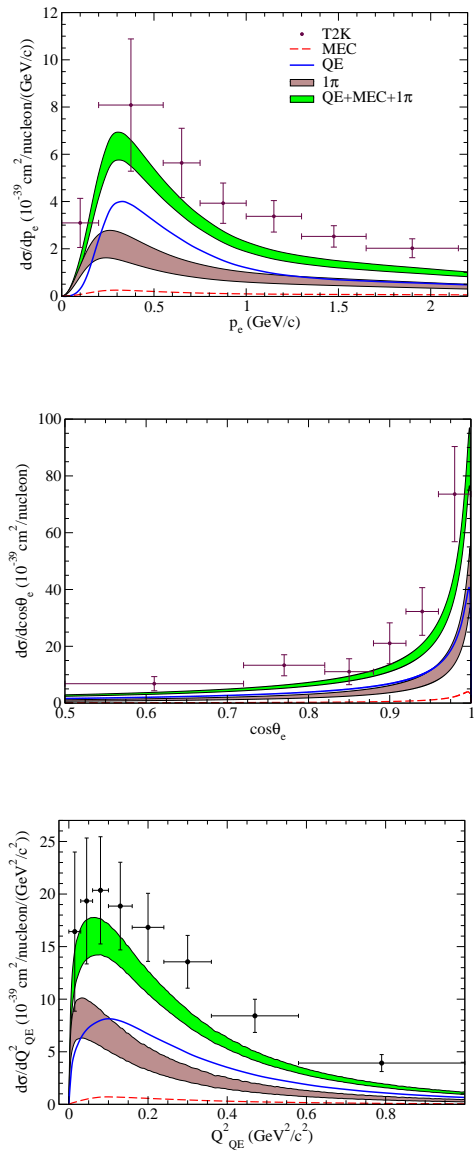

  \includegraphics[scale=0.25]{T2K_electron_Pe.eps} \\
\vspace{1.cm}
\includegraphics[scale=0.25]{T2K_electron_ang.eps} \\
\vspace{1.cm}
  \includegraphics[scale=0.25]{T2K_electron_Q2.eps}
\vspace{1.cm}
\caption{(Color online) The  CC-inclusive T2K flux-folded $\nu_e$-$^{12}$C differential cross section per nucleon evaluated in the SuSAv2 model is displayed as a function of the electron momentum (top), $\cos\theta_e$ (middle) and $Q^2_\text{QE}$ (bottom). The separate contributions of the QE, 1$\pi$ and vector 2p2h MEC are displayed. The data are from \cite{T2Kincl_electron}. }
\label{fig:T2Kelectron}
\end{figure*}

\begin{figure*}[ht]
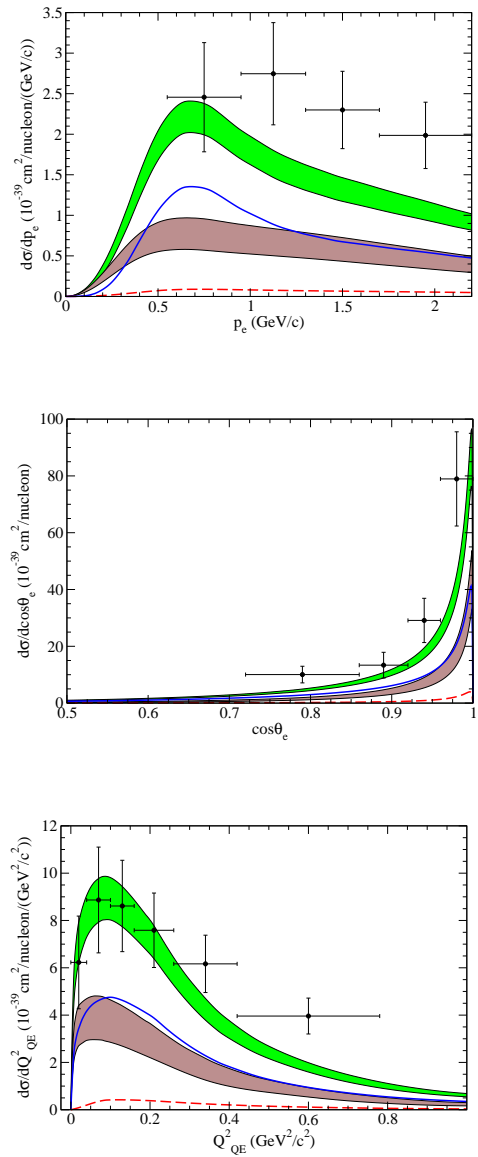

  \includegraphics[scale=0.25]{T2K_electron_Pe_cut.eps} \\
\vspace{1.cm}
\includegraphics[scale=0.25]{T2K_electron_ang_cut.eps} \\
\vspace{1.cm}
  \includegraphics[scale=0.25]{T2K_electron_Q2_cut.eps}
\vspace{1.cm}
\caption{(Color online) As for Fig.~\ref{fig:T2Kelectron}, but now only for electrons with $p_e>$ 0.55 GeV/c and $\cos\theta_e>$ 0.72. }
\label{fig:T2Kelectroncut}
\end{figure*}

\subsection{T2K}\label{T2K}

In Fig.~\ref{fig:T2K} we show the CC-inclusive $\nu_\mu-^{12}$C double-differential cross section per nucleon versus the muon momentum, $p_\mu$, for different angular bins, folded with the T2K flux. The QE curve corresponds to the SuSAv2 model illustrated in Sect.~\ref{model} (see~\cite{Gonzalez-Jimenez:2014eqa} for details). The resonant pion production curve ($1\pi$) is obtained with the non-QE scaling function described above. As in the previous case of electron scattering, the band corresponds to the theoretical uncertainty associated with the extraction of the non-QE scaling  function discussed in the previous section. The MEC curve corresponds to the fully relativistic calculation of 2p2h excitations induced by pionic vector two-body current of \cite{De Pace:2003xu} and parameterized in~\cite{Megias:2014qva}.

We observe that the model yields excellent agreement with the data. Moreover, the main contribution in the cross section comes from the QE and pion production mechanisms. On the contrary, MEC play a minor role at these kinematics, a result that is somehow different from the one found in~\cite{Martini:2014dqa}.
It should be noted however that the two calculations differ in various respects: first, the present model does not include the axial two-body current, as explained in the previous section; second, the two calculations, although in principle similar, involve different approximations in the way they account for relativistic effects -- the calculation of~\cite{De Pace:2003xu} being exactly relativistic -- and in some important technical details in the multidimensional integration leading to the results (see~\cite{Simo:2014wka,Simo:2014esa}).
Indeed, the MEC contributions here are so small that, even were AA and VA contributions that are as large as these VV contributions to be included, the net effect would still not be very significant.

As shown in the analysis of the non-QE scaling function (Fig.~\ref{fig:fDelta}), scaling is not fulfilled at $\psi_\Delta\gtrsim0-0.5$
due to other inelastic processes whose contributions start to be more significant at high kinematics (\emph{i.e.} high momentum
transferred). However, contributions beyond $\psi_\Delta=0.5$
are not very significant at the kinematics involved in the $\nu_{\mu}$ T2K experiment, as can be seen in Fig.~\ref{fig:T2K_cuts}.
Indeed, the effects in the 1$\pi$ cross sections associated with this positive-$\psi_\Delta$ tail (high values of the transfer energy)
are less than $10$--$12$\%.
This supports the reliability of our model to be applied to the description of T2K muon-neutrino data where the average value
of the neutrino energy is $\langle E_{\nu_\mu}\rangle \sim 0.85$~GeV. Furthermore, upon evaluating the importance of high momentum transfer contributions by cutting the predictions at various values of $q$, we have confirmed that for muon momenta at T2K kinematics the model appears to be robust.

In Fig.~\ref{fig:T2Kelectron} we compare
our predictions with recent T2K data corresponding to electron-neutrino scattering~\cite{T2Kincl_electron}. We show results for
the flux-folded $\nu_e$-$^{12}$C differential cross section against the electron momentum (top panel), scattering angle (middle)
and QE transferred four-momentum (bottom). As observed, the model understimates the data at high $p_e$ and $Q^2_\text{QE}$, in contrast with the situation observed
in the previous case, {\it i.e.,} muonic neutrinos. This clearly indicates that high-inelasticity processes, which are not incorporated
in our formalism, have a significant contribution in the analysis of this experiment. Moreover, this is also consistent with the
electronic neutrino flux with an average energy $\langle E_{\nu_e} \rangle \sim 1.3$~GeV, which is significantly larger than the value
corresponding to the case of muonic neutrinos, $\langle E_{\nu_\mu} \rangle \sim 0.85$~GeV. This also reflects the much more important
tail in the electronic neutrino flux that extends to very high neutrino energies. From analyses of $q$ and $\psi_{\Delta}$ cuts as discussed above we find that effects from contributions lying above the $\Delta$ might provide 20-30\% more strength in some cases.  In particular, looking at the lowest panel in Fig.~5, one finds that the cuts have very little effect near the peak of the cross section at low $Q^2_{QE}$, but are more significant in the tail ($\sim$30\%). Thus this shortfall in the latter region is probably not unexpected.

For completeness, we also show in Fig.~\ref{fig:T2Kelectroncut}
the ``reduced phase-space'' results for the inclusive $\nu_e-^{12}$C reaction where a better
agreement with low-$Q^2_{QE}$ data is reached when considering only very forward processes ($\cos\theta_e>0.72$) and rejecting
low values for the electron momentum ($p_e>0.55$ GeV/c).

\begin{figure}[ht]
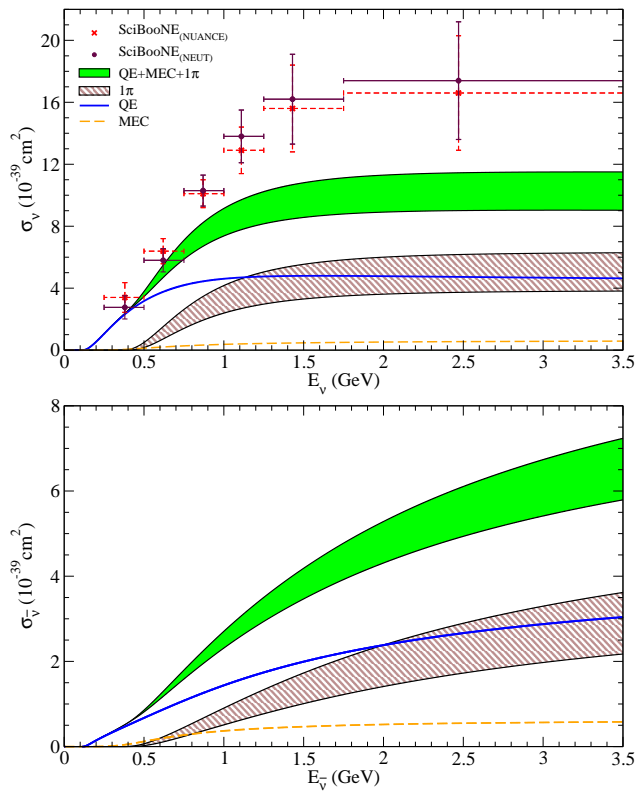

 \includegraphics[scale=0.3,angle=0]{SciBooNE.eps} \\
 \includegraphics[scale=0.3,angle=0]{SciBooNE_bar.eps}
\caption{(Color online) The CC-inclusive $\nu_\mu$ cross section on a polystyrene target (C$_8$H$_8$)  per nucleon evaluated in the SuSA model as a function of the neutrino energy. The SciBooNE data are from \cite{SciBooNEincl}. }
\label{fig:SciBooNE}
\end{figure}

\subsection{SciBooNE}\label{SciBooNE}

In this section we apply our model to the kinematics involved in the SciBooNE experiment~\cite{SciBooNEincl}, that
corresponds to CC inclusive $\nu_\mu$ scattering on a polystyrene target (C$_8$H$_8$). In this case the data are presented as a
total unfolded integrated cross section as a function of the neutrino energy.
In the unfolding procedure the neutrino energy is reconstructed from the kinematics of the outgoing lepton assuming that the process
is purely quasielastic.
This is a strong assumption, especially at high neutrino energies where inelastic processes become important, and makes the
comparison of theoretical results with data very delicate. We should notice that the averaged value of the neutrino energy for
SciBooNE is 0.76 GeV whereas the unfolding is extended up to 5 GeV.
This makes an important difference with T2K data, which are not unfolded, and the corresponding calculations, which
involve an integration over the neutrino flux.

Our results for $\nu_\mu$ inclusive scattering are compared with SciBooNE data  in the top panel of Fig.~\ref{fig:SciBooNE}.
We also show for completeness predictions corresponding to antineutrinos (bottom panel).

In the case of neutrinos (top panel), the QE contribution grows with the neutrino energy up to $E_{\nu_\mu}\sim 1$ GeV, where it
saturates to a value of the cross section close to $\sim\!\! 5\times 10^{-39}$ cm$^2$. On the other hand, the resonant pion production
result, $1\pi$, (displayed as the brown band) starts its contribution at $E_{\nu_\mu}$ slightly below $0.5$~GeV, and reaches its saturation
value $\sim\!\! 5 \times 10^{-39}$ cm$^2$ at $E_\nu\sim 1.5$~GeV. The sum of the two contributions agrees with the two lowest experimental
points taken at $E_{\nu_\mu}\approx0.4$ and $0.65$~GeV, respectively.
On the contrary, at higher neutrino energies the model clearly underpredicts the data by approximately a factor two at $E_{\nu_\mu}$
above $1.5$~GeV. This result clearly indicates that new channels and higher resonances should be included in the model
(see also comments in the case of T2K $\nu_e$ data).
This is also consistent with previous calculations based on RPA~\cite{Nieves:2011pp,Martini:2014dqa}
that were restricted to neutrino energies below $E_{\nu_\mu}\sim 1.2$~GeV.
Although not shown in the figures, we have evaluated the contribution in the resonant pion production ascribed to the
kinematical region above $\psi_\Delta=0.5$, {\it i.e.,} the tail where high inelasticities might give significant effects:
in this case this region ($\psi_\Delta\geq 0.5$) provides a contribution of the order of $\sim 30\%$ of the total integrated
cross section at high energies (where saturation is already reached).

Finally, as for the T2K case discussed in the previous section, we find the pionic 2p2h MEC to give a minor contribution, at most of the order of $\sim$7\% of the total cross section at the highest energies. The uncertainty introduced by the use of different parametrizations of the axial $N\to\Delta$ form factors is very small (less than $5.5\%$ of the total cross section at the highest energies).

In the case of antineutrinos (bottom panel) we observe that MEC contributions are about 9\% at the highest energies. Another important difference between neutrinos and antineutrinos concerns the property of saturation. Whereas the neutrino cross section already saturates at energies of the order of $E_{\nu_\mu}\sim 1.5$~GeV, and this result applies to both QE and pionic channels, the antineutrino cross section continues to grow for increasing values of $E_{{\bar\nu}_\mu}$. Results in Fig.~\ref{fig:SciBooNE} are also consistent with RPA predictions in~\cite{Nieves:2011pp,Martini:2014dqa}.
The analysis of T2K results on antineutrino CC inclusive cross sections (not completed yet) will undoubtedly help us in disentangling the
specific roles played by the different ingredients that enter in the description of the scattering reaction.

\section{Conclusions}\label{Conclusions}

We have compared the predictions of the recently revised superscaling model (SuSAv2), devised for QE scattering and extended to the $\Delta$-resonance production region, with the available inclusive data for charged current muon (electron) neutrino-$^{12}$C reactions of the T2K and SciBooNE experiments, where the mean neutrino energy is $0.85$~GeV ($1.3$~GeV) and $0.8$~GeV, respectively. The model also includes 2p2h excitations induced by vector meson-exchange currents carried by the pion and has been tested against inclusive electron scattering. Moreover, the model is fully relativistic and can therefore be applied at high energies, provided the relevant physics mechanisms are taken into account.

Our main conclusions can be summarized as follows:
\begin{enumerate}
\item The present approach provides a very good representation of the T2K $\nu_\mu$ experimental data. A similar comment applies to the SciBooNE data for neutrino energy below $0.7-0.8$~GeV.

\item On the contrary, the model fails in reproducing SciBooNE data at higher neutrino energies, as well as T2K $\nu_e$ data. This is a clear signal of the relevance of other reaction mechanisms (not included in the model yet) such as resonances beyond the $\Delta$, multi-meson production and deep inelastic scattering. Work is in progress to implement these processes in the model.

\item Pionic (vector) meson-exchange currents in neutrino scattering are shown to play a minor role ($<$10\%) for all of the kinematical situations considered here. Axial-vector MEC contributions have yet to be included.

\item The uncertainty related to the poorly known axial form factors entering in the $N\to\Delta$ current does not present a significant impact ($<$6\%) at the experimental kinematics analyzed.

\end{enumerate}

\begin{acknowledgments}

 This work was partially supported by INFN under project MANYBODY, by Spanish DGI and FEDER funds (FIS2011-28738-C02-01, FPA2013-41267), by the Junta de Andalucia, by the Spanish Consolider-Ingenio 2000 program CPAN (CSD2007-00042), by the Campus of Excellence International (CEI) of Moncloa project (Madrid) and Andalucia Tech, by the Bulgarian National Science Fund under contracts No. DFNI-T02/19 and DFNI-E02/6 (M.I.), by the Office of Nuclear Physics of the U.S. Department of Energy under Grant Contract Number DE-FG02-94ER40818 (T.W.D.) and by the 7th European Community Framework Programme Marie Curie IOF ELECTROWEAK (O.M.).
R.G.J. acknowledges financial help from the Interuniversity Attraction Poles Programme initiated by the Belgian Science Policy Office.
The authors would like to thank Marco Martini for interesting discussions.
\end{acknowledgments}

\end{document}